\documentclass[epjCONF]{svjour}
\usepackage{graphics}
\usepackage[varg]{txfonts} 
\usepackage[latin1]{inputenc}
\session-title{MESON2012 - 12th International Workshop on Meson Production, Properties and Interaction} 
\begin{document}
\title{Exclusive meson pair production in proton-proton collisions}
\author{Piotr Lebiedowicz\inst{1}\fnmsep\thanks{\email{Piotr.Lebiedowicz@ifj.edu.pl}} 
\and Antoni Szczurek\inst{2} }
\institute{Institute of Nuclear Physics PAN, PL-31-342 Cracow, Poland 
\and Institute of Nuclear Physics PAN, PL-31-342 Cracow, Poland, and\\
University of Rzesz\'ow, PL-35-959 Rzesz\'ow, Poland}
\abstract{
We present a study of the exclusive production of meson pairs
in the four-body $pp \to pp \pi^{+} \pi^{-}$, $pp K^{+}K^{-}$ 
reactions at high energies which constitute
an irreducible background to resonance states
(e.g. $\phi$, $f_{2}(1270)$, $f_{0}(1500)$, $f_{2}'(1525)$, $\chi_{c0}$).
We consider central diffractive contribution
mediated by Pomeron and Reggeon exchanges 
and new diffractive mechanism of emission of pions/kaons from the proton lines.
We include absorption effects due to proton-proton interaction 
and pion/kaon rescattering.
Predictions for the total cross section and differential
distributions in pion/kaon rapidity and transverse momentum 
as well as two-pion/kaon invariant mass are presented 
for the RHIC, Tevatron and LHC colliders. 
Finally we consider a measurement of exclusive production
of a scalar $\chi_{c0}$ meson via $\chi_{c0} \to \pi^{+} \pi^{-}$, $K^{+} K^{-}$ decay. 
}
\maketitle
\section{Introduction}
\label{intro}
Central exclusive production (CEP) processes of the type
$pp \to p X p$, where $X$ represents
the centrally produced state separated from the two very forward protons
by large rapidity gaps,
significantly extend the physics programme at hadron colliders
\cite{COMPASS,Turnau,Zurek}.
As discussed in \cite{LPS11,Lang_Khoze}, the measurement of $\chi_{c0}$
CEP via two-body decay channels to light
mesons is of special interest for both studying the dynamics
of heavy quarkonia and for testing the QCD framework of CEP.
However, in this case we may expect a sizeable background resulting
from meson pair production; such a non-resonant 
contribution should therefore be carefully evaluated.
We have studied the four-body 
$pp \to pp \pi^{+} \pi^{-}$ \cite{SL09,LSK09,LS10}, $pp K^{+}K^{-}$ \cite{LS11_kaons}
reactions which constitute an irreducible background to resonance states.

CEP processes have been successfully observed at the Tevatron \cite{Zurek}
by selecting events with large rapidity gaps.
However, the CDF measurement \cite{CDF} of $\chi_{c}$
with no other hadrons in the final state,
in the $\chi_c \to J/\psi + \gamma$ channel 
does not allow a separation of the different $\chi_{cJ}$ states.
It may be possible to isolate the $\chi_{c}$ CEP contribution 
via two-body decay channels, especially in 
$\chi_{cJ} \to \pi^{+} \pi^{-}$ \cite{LPS11}, 
$K^{+} K^{-}$ \cite{LS11_kaons} and $p \bar{p}$ decay.
We recall that such hadronic channels are ideally
suited for spin-parity analysis of the $\chi_{c}$ states:
in particular the fact that
the branching fraction to these channels 
are relatively larger for scalar meson
than for the tensor meson and even absent for the axial meson.
A much smaller cross section for $\chi_{c2}$ production as obtained
from theoretical calculation means that only $\chi_{c0}$ 
will contribute to the signal \cite{PST_chic,LKRS10}.

A new area of experimental studies of CEP with
tagged forward protons and obtained the invariant mass of the
pion pair using tracks reconstructed in the STAR Time Projection Chamber
(TPC) has just started \cite{Turnau}.

At the LHC CEP studies looks also very promising 
both in the low dipion mass regime and at high masses.
In Ref. \cite{SLTCS11} a possible measurement of the exclusive $\pi^+ \pi^-$ production
at the LHC with tagged forward protons has been studied.
The $pp \to nn \pi^+ \pi^+$ process \cite{LS11} is also interesting
for possible future experiments at high energies.

\section{Formalism}
\label{formalism}

\begin{figure}[!h]
\centering
\resizebox{0.8\columnwidth}{!}{%
\includegraphics{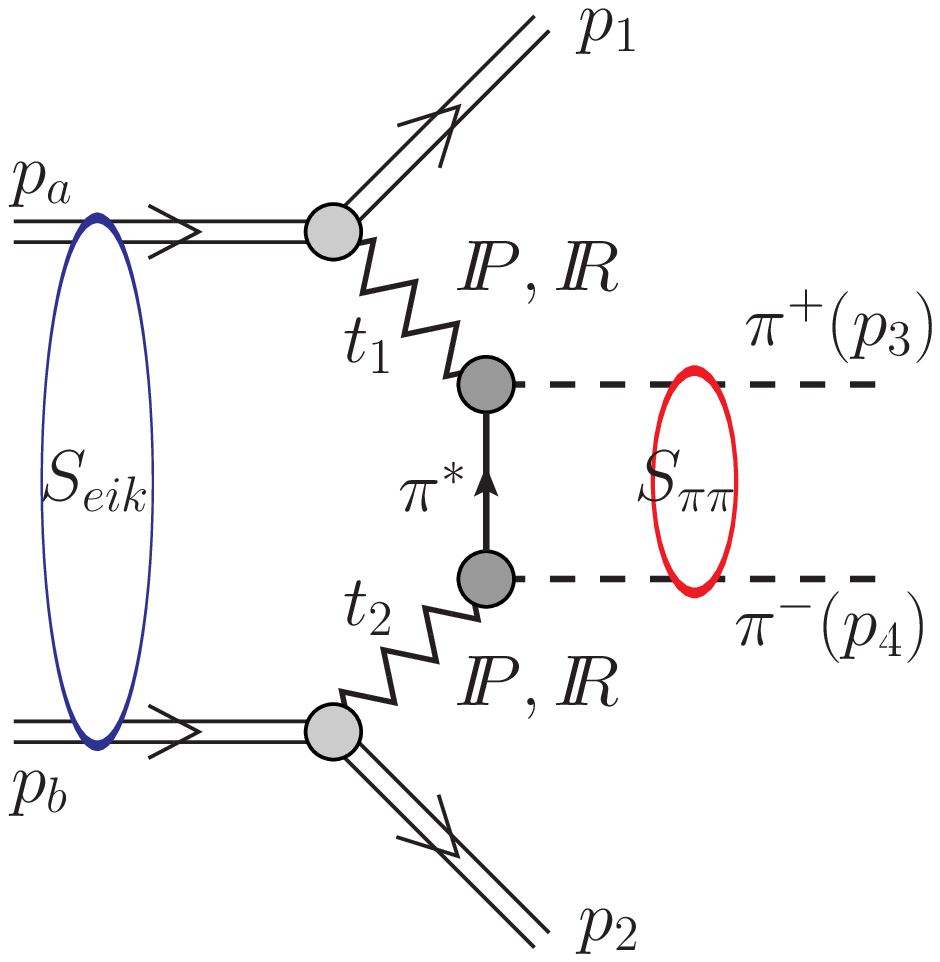}
\includegraphics{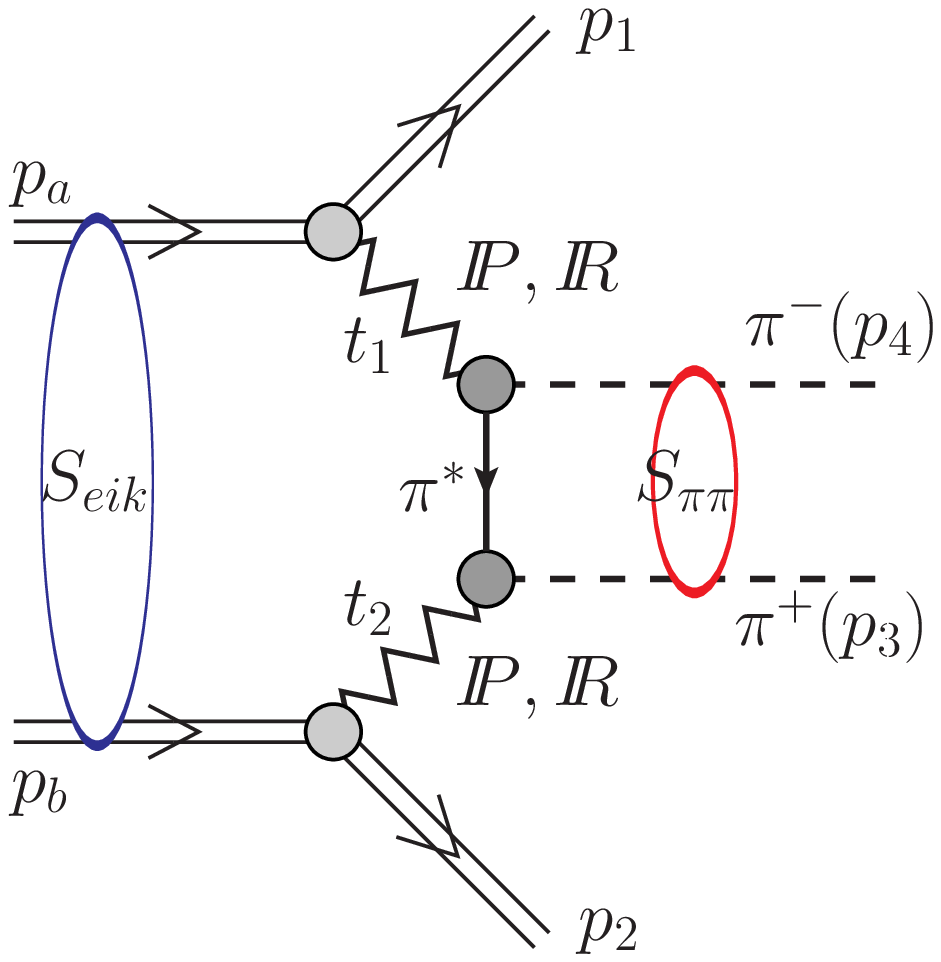} \qquad \qquad \qquad \qquad
\includegraphics{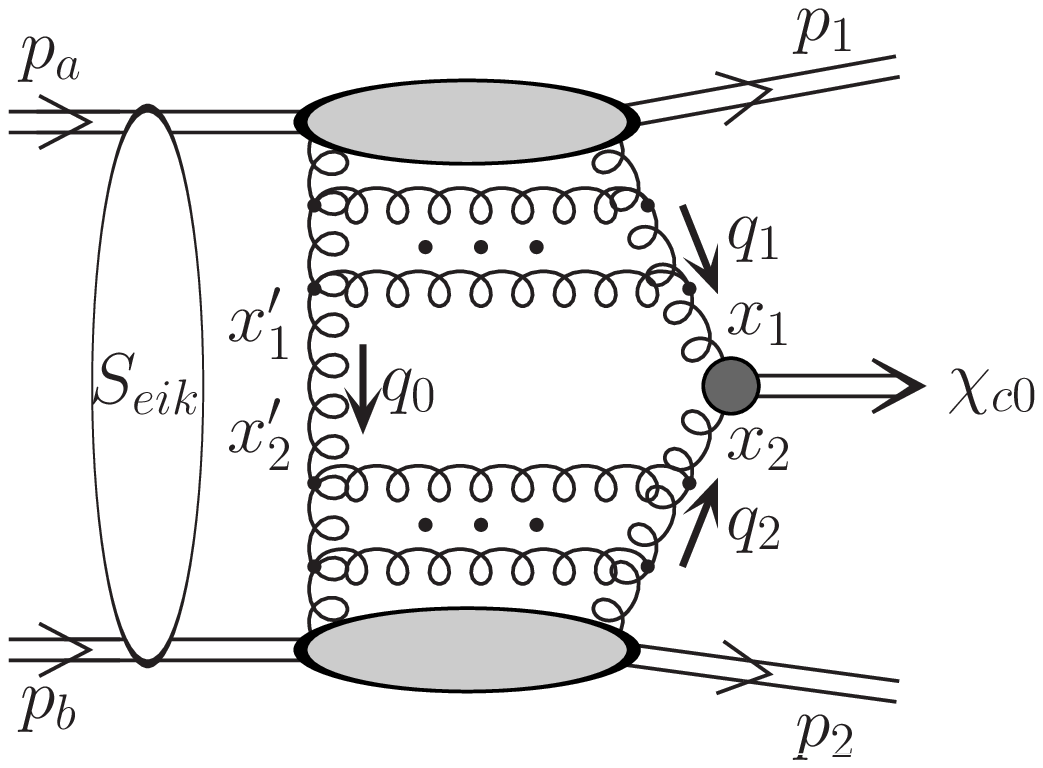} }
\caption{
Left panel: The central diffractive mechanism of exclusive $\pi\pi$ or $KK$ pair production
including the absorptive corrections due to proton-proton interactions
and pion/kaon rescattering.
Right panel: The QCD mechanism of $\chi_{c0}$ CEP.}
\label{fig:1}
\end{figure}

The dominant mechanism of the exclusive production of
light meson pairs at high energies is sketched in Fig.\ref{fig:1} (left panel).
The formalism used to calculate
of non-resonant background amplitude is explained in detail in Refs \cite{LS10,LS11_kaons}.
The Regge parametrization of the scattering amplitude includes 
both Pomeron and subleading Reggeon exchanges.
Our model with the parameters taken from the Donnachie-Landshoff analysis
of the total $\pi N$ or $KN$ cross sections
sufficiently well describes the elastic data for energy $\sqrt{s} > 3$ GeV.
The form factors correcting for the off-shellness of the intermediate pions/kaons
are parametrized as
$F_{\pi/K}(\hat{t}/\hat{u})=
\exp\left(\frac{\hat{t}/\hat{u}-m_{\pi/K}^{2}}{\Lambda^{2}_{off}}\right)$,
where the parameter $\Lambda_{off}^{2} = 2$ GeV$^{2}$ is obtained from a fit 
to the ISR experimental data \cite{ABCDHW89,ABCDHW90}.

The QCD amplitude for exclusive central diffractive $\chi_{c0}$ meson production,
sketched in Fig.\ref{fig:1} (right panel),
was calculated within the $k_{t}$-factorization approach
including virtualities of active gluons \cite{PST_chic}
and the corresponding cross section is calculated with the help of
unintegrated gluon distribution functions (UGDFs).
In Ref.\cite{LPS11} we have performed detailed studies of several differential distributions of $\chi_{c0}$ meson production.

\section{Results}
\label{results}

\begin{figure}
\resizebox{1.\columnwidth}{!}{%
  \includegraphics{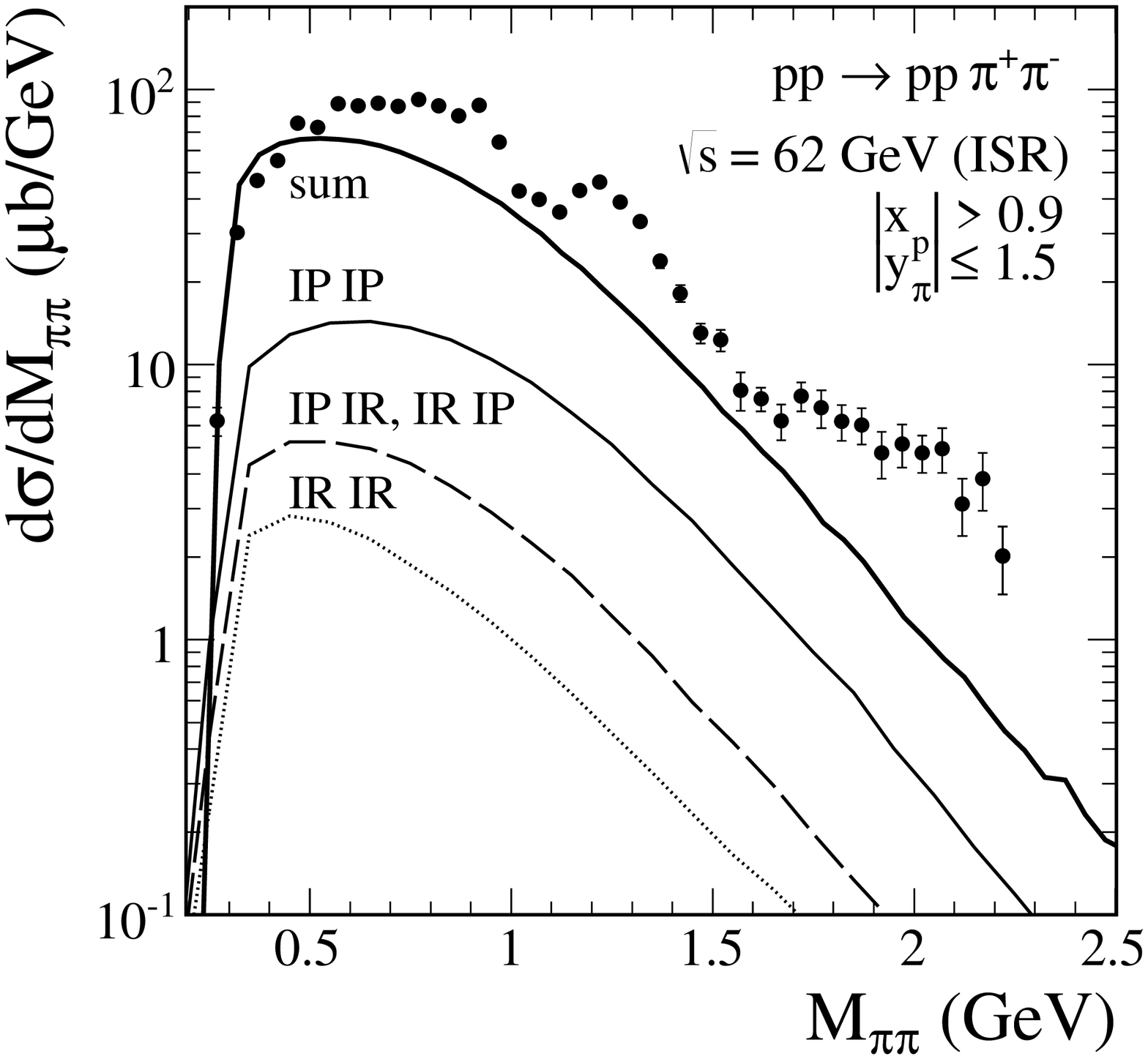} 
  \includegraphics{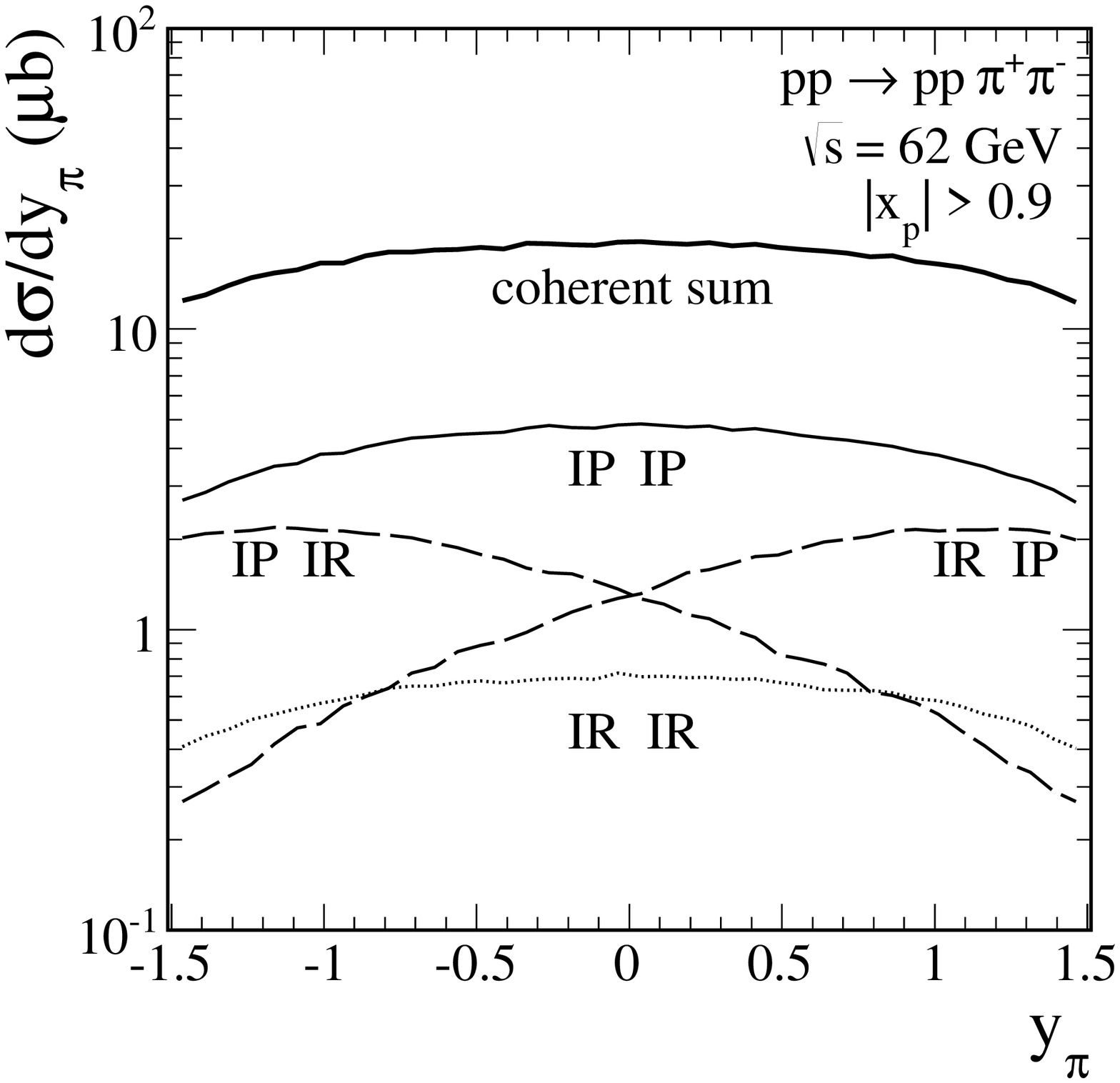} 
  \includegraphics{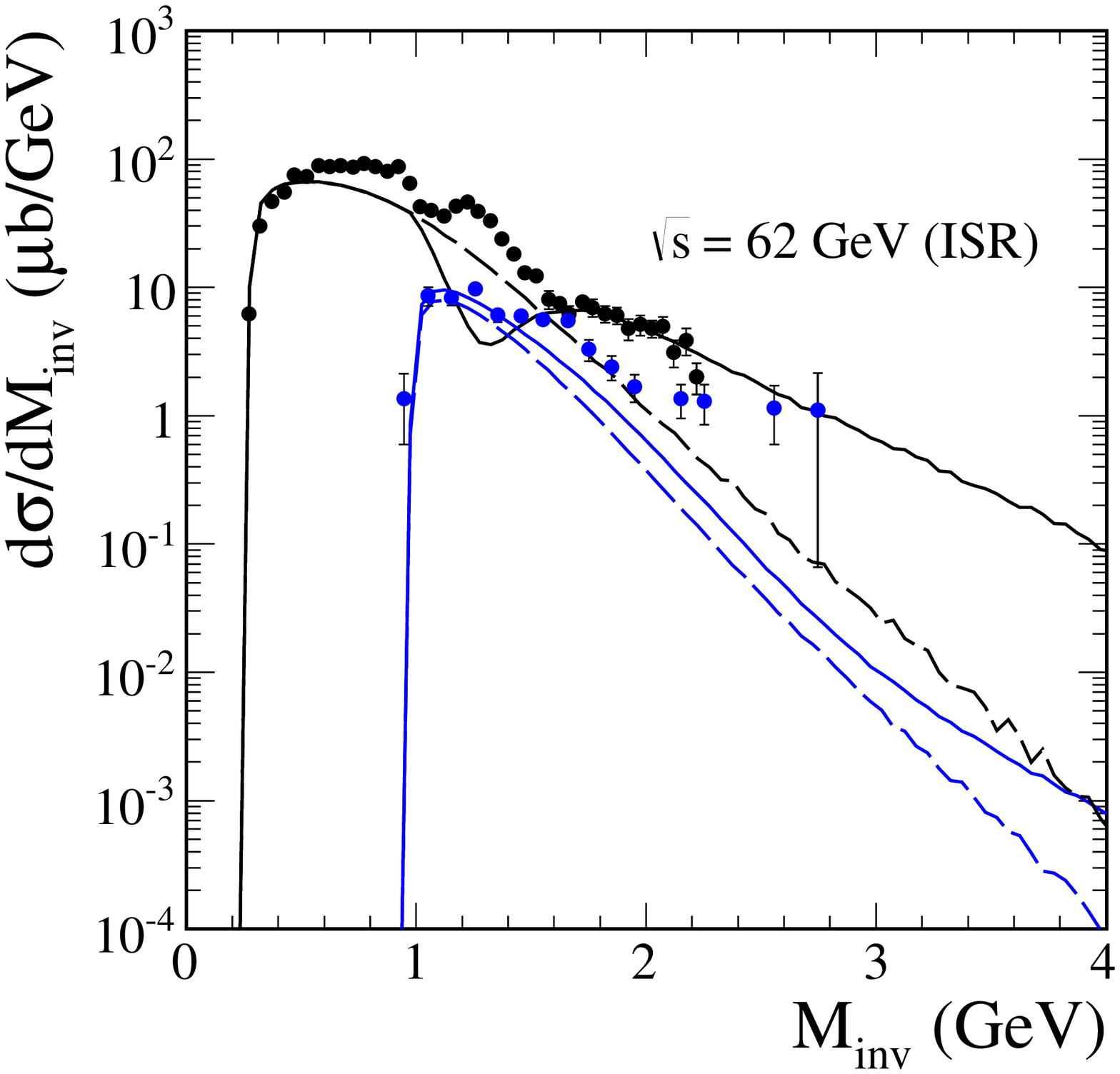} }
\caption{Left panel:
Differential cross section $d\sigma/dM_{\pi\pi}$
at $\sqrt{s} = 62$ GeV with experimental cuts 
relevant for the ISR data \cite{ABCDHW90}.
The different lines correspond to the situation when all 
and only some components of the Pomeron and Reggeon exchanges 
in the amplitude are included 
(Pomeron-Pomeron component dominates at midrapidities of pions 
and Pomeron-Reggeon (Reggeon-Pomeron)
peaks at backward (forward) pion rapidities, respectively.
Central panel:
Differential distributions in pion rapidity.
The camel-like shape of the distributions 
is due to the interference of components in the amplitude.
Right panel: 
Comparison of two-pion/kaon invariant mass distribution 
(black/blue lines, respectively),
without (solid lines) and with (dashed lines)
rescattering effects in final state.
The absorption effects due to $pp$-interaction
were included in the calculations.
}
\label{fig:2}       
\end{figure}

\begin{figure}
\resizebox{1.\columnwidth}{!}{%
\includegraphics{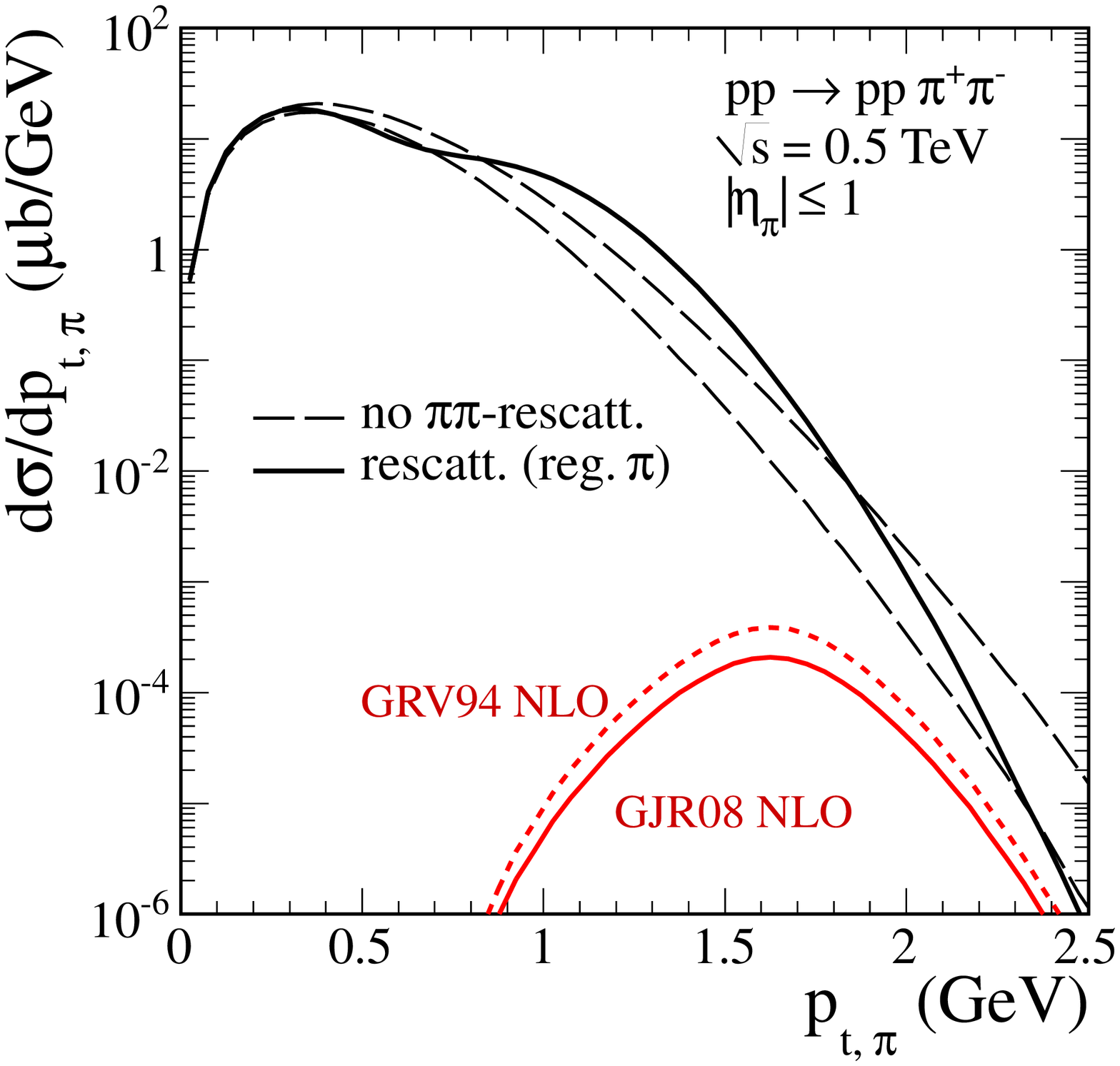}
\includegraphics{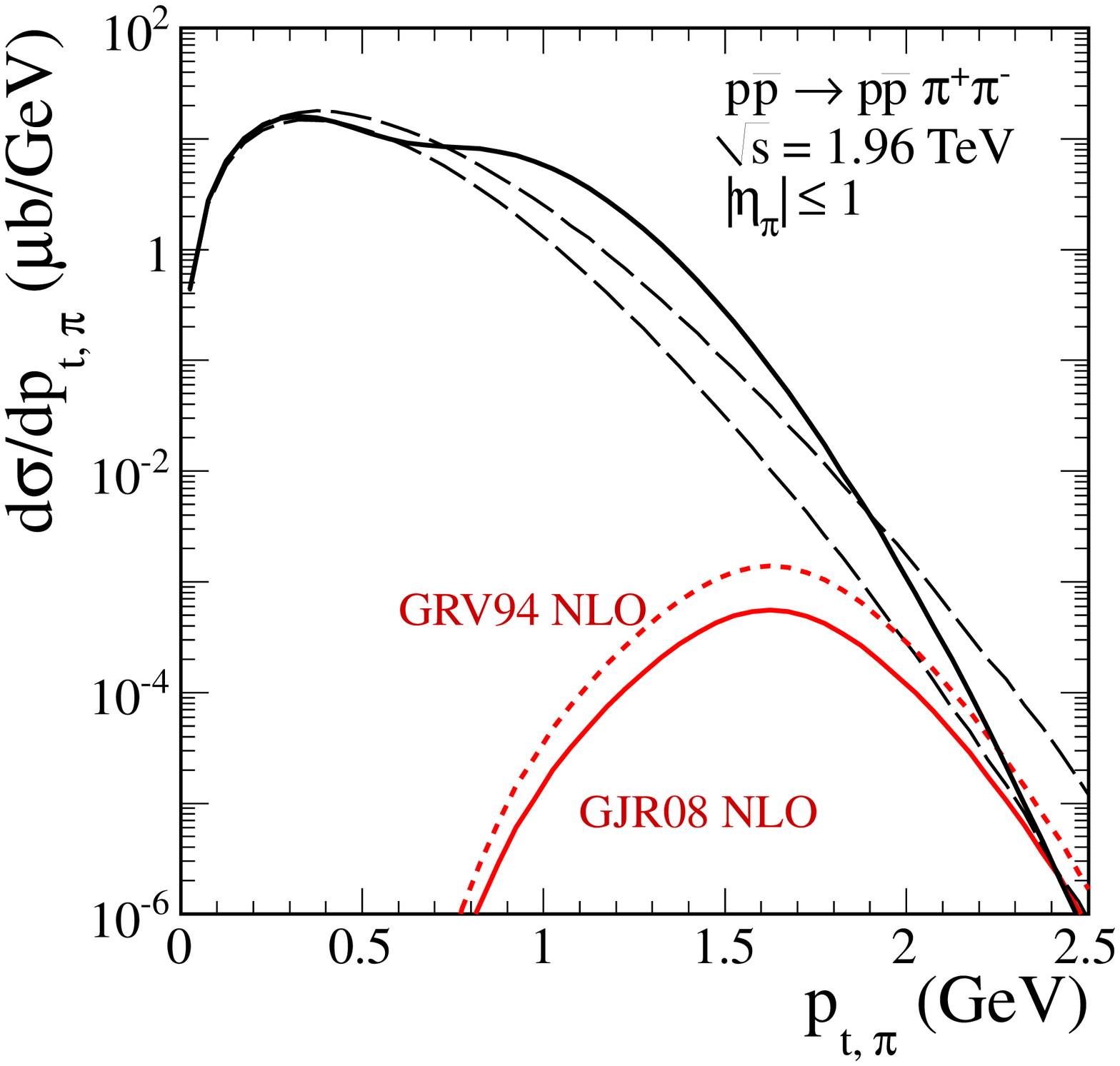}
\includegraphics{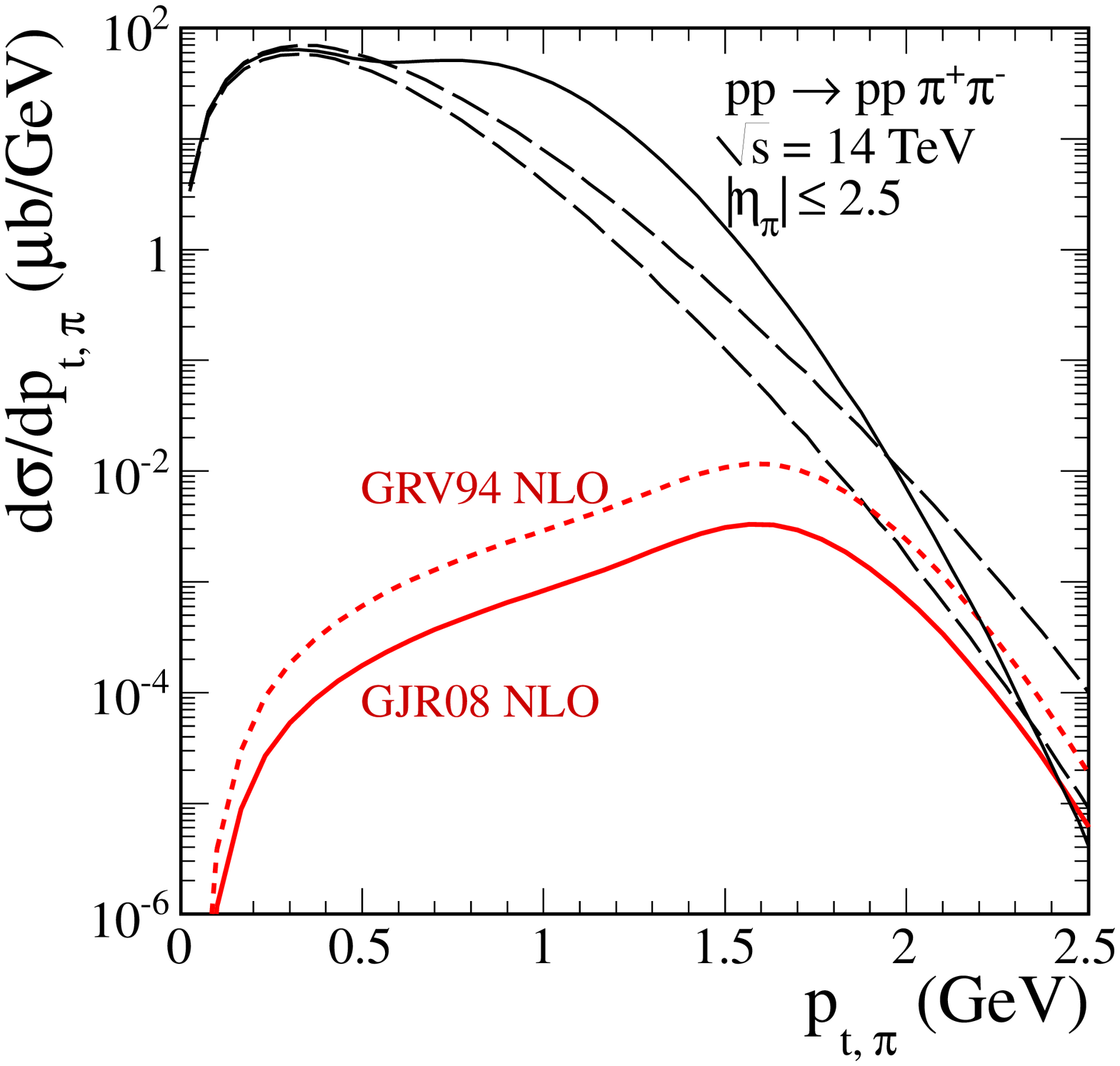} }
\resizebox{1.\columnwidth}{!}{%
\includegraphics{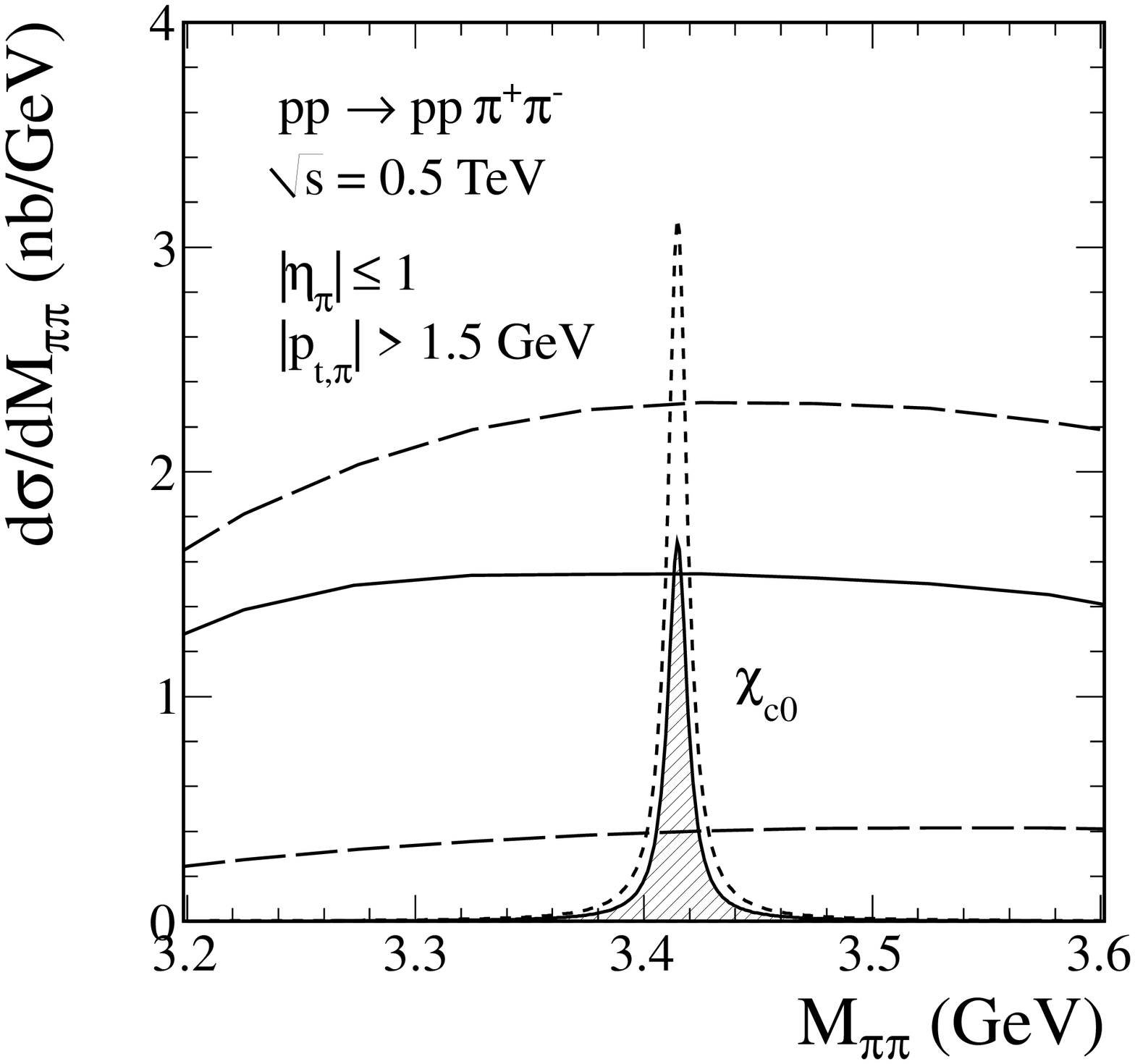}
\includegraphics{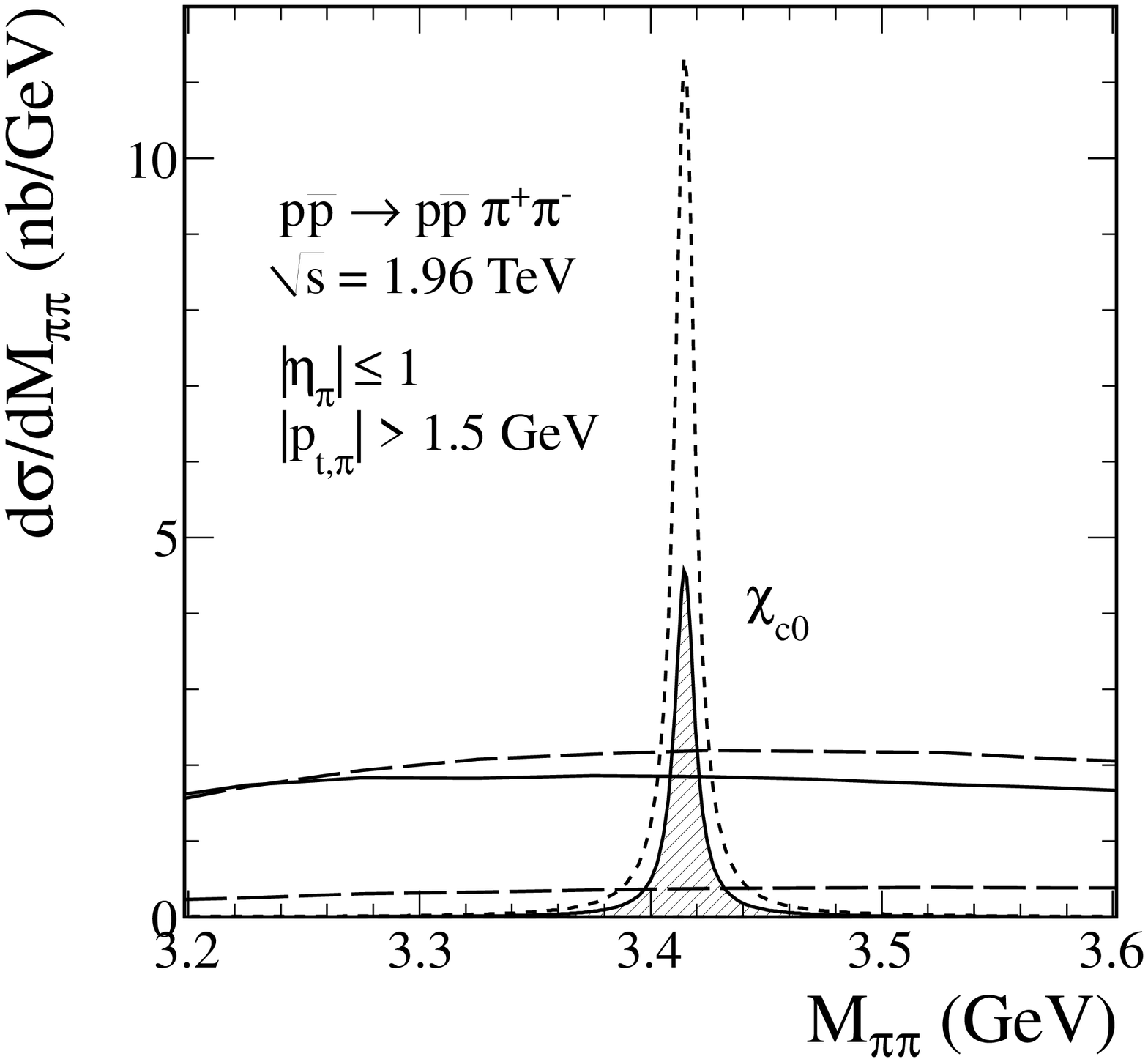}
\includegraphics{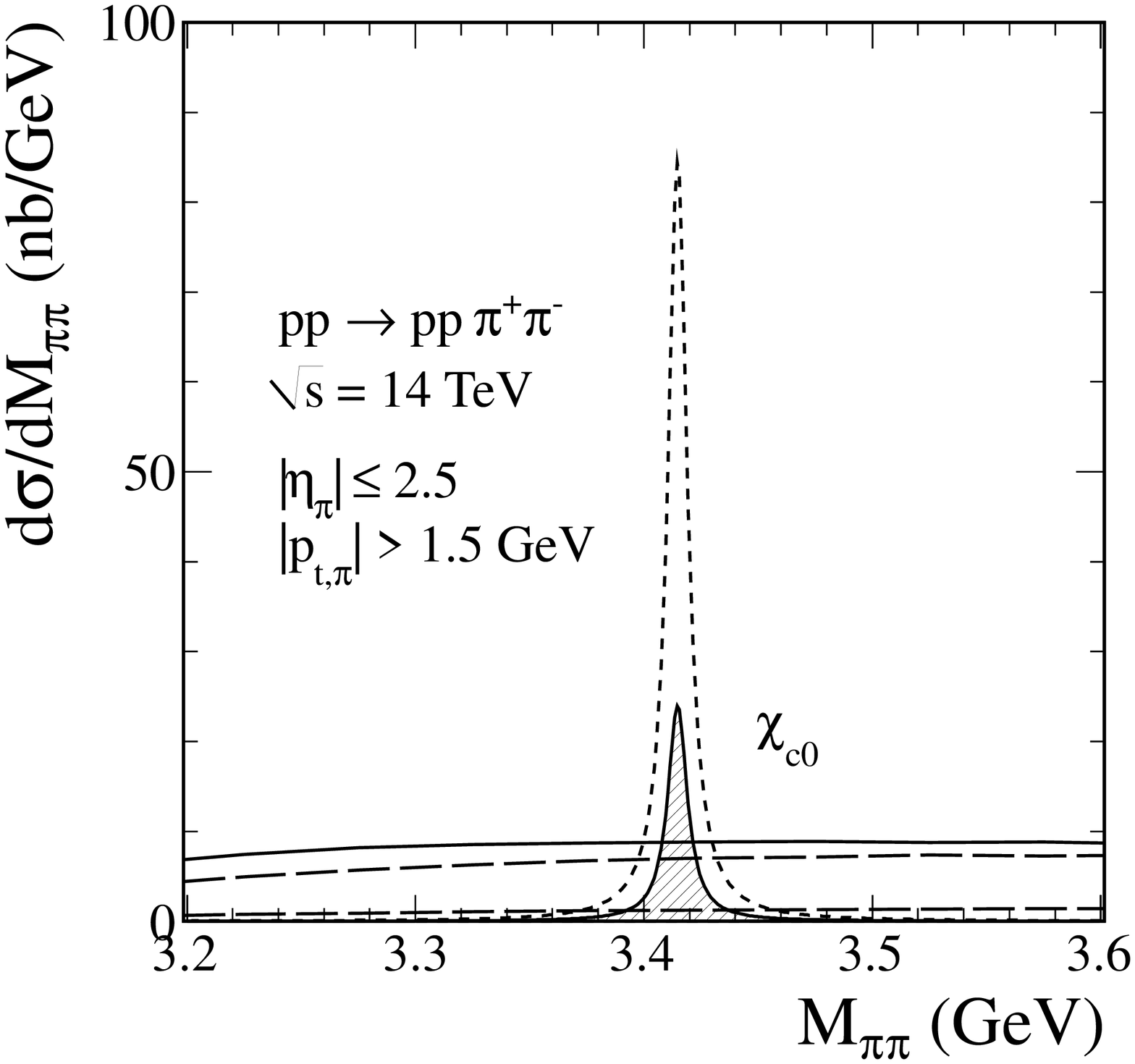} }
\caption{Upper panel: 
Differential cross section $d\sigma/dp_{t,\pi}$ at $\sqrt{s} = 0.5,
1.96, 14$ TeV with cuts on the pion pseudorapidities.
Results for the $\pi\pi$ continuum with the meson propagator
and with the cut-off parameter $\Lambda_{off}^{2}$ = 1.6, 2 GeV$^{2}$ 
(lower and upper dashed lines, respectively) as well as
with the generalized pion propagator and $\pi\pi$-rescattering (solid line) 
are presented.
The absorption effects both for the signal and background 
were included in the calculations.
Bottom panel: The $\pi^{+}\pi^{-}$ invariant mass distribution
with the relevant restrictions in the pion
pseudorapidities and pion transverse momenta.
In the calculation of the $\chi_{c0}$ distributions we have used 
two choices of collinear gluon distributions: 
GRV94 NLO (dotted lines) and GJR08 NLO (filled areas).
}
\label{fig:3}       
\end{figure}

In Fig.~\ref{fig:2} we compare our results with
CERN ISR experimental data \cite{ABCDHW89,ABCDHW90} at $\sqrt{s} = 62$ GeV.
The mass spectra shows strong resonances structure 
attributed to $f_{0}$ and $f_{2}$ states; their features change with transverse momentum.
We see distributions in two-pion invariant mass and in pion rapidity
when all (solid line) and only some components in the amplitude are included.
Distributions in pion transverse momenta Fig.~\ref{fig:3} (upper panel)
and in $M_{\pi\pi}$ Fig.~\ref{fig:3} (bottom panel)
both for the signal ($\chi_{c0}$) and background are presented.
The pions from the $\chi_{c0}$ decay are placed at larger $p_{t,\pi}$. 
This can be therefore used to improve the signal-to-background ratio 
by imposing an additional cuts on both pion transverse momenta
$|p_{t,\pi}| > 1.5$ GeV.
Measurements of other decay channels, e.g. $K^{+}K^{-}$,
are possible as well.

\begin{figure}
\resizebox{0.3\columnwidth}{!}{%
  \includegraphics{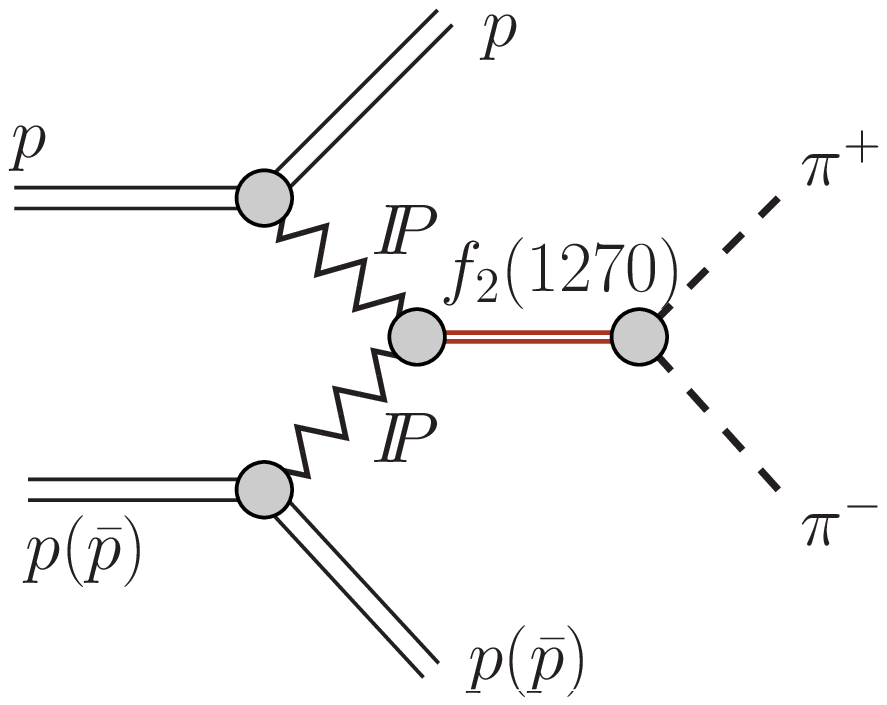} }
\resizebox{0.7\columnwidth}{!}{%
  \includegraphics{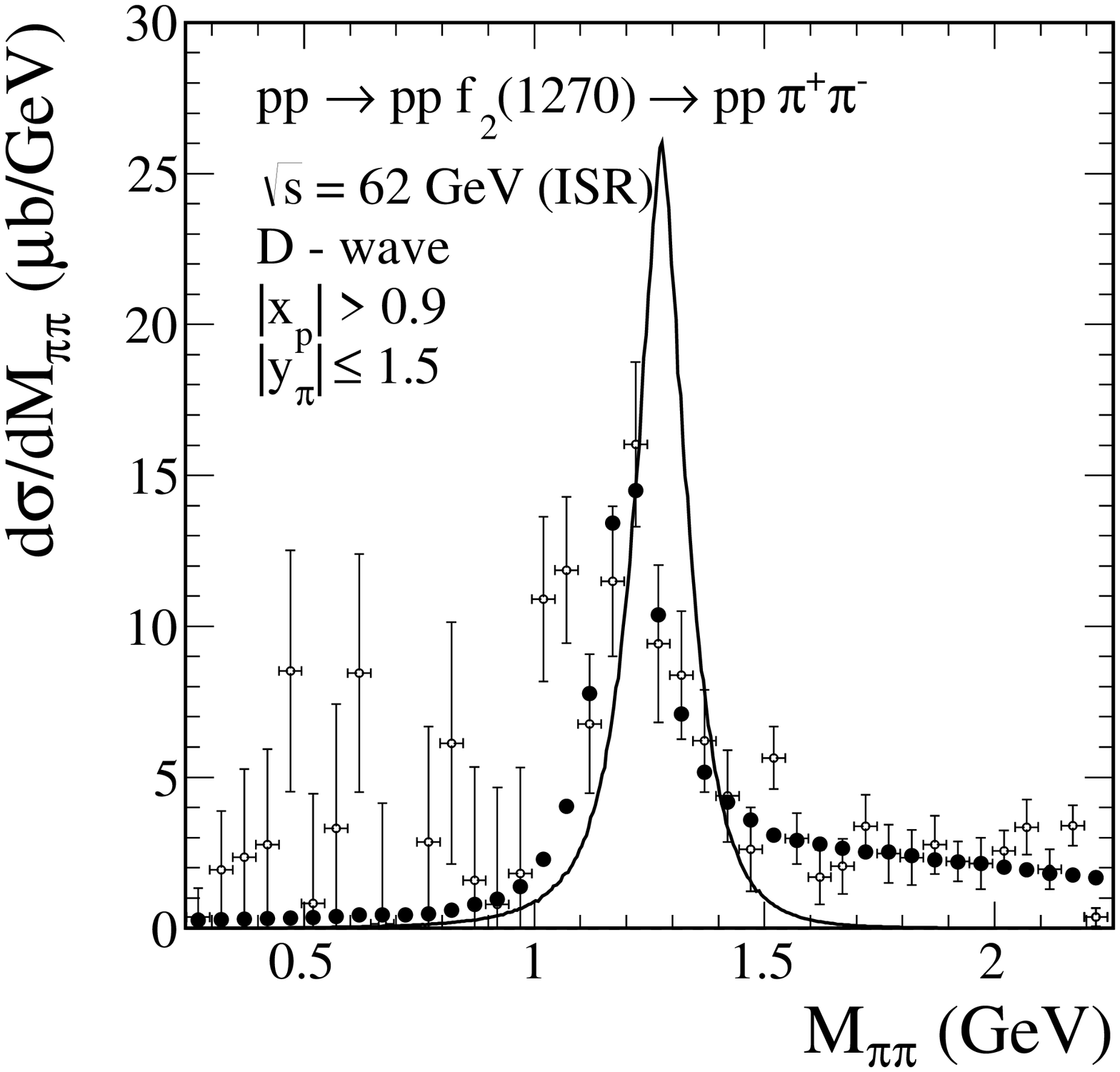} 
  \includegraphics{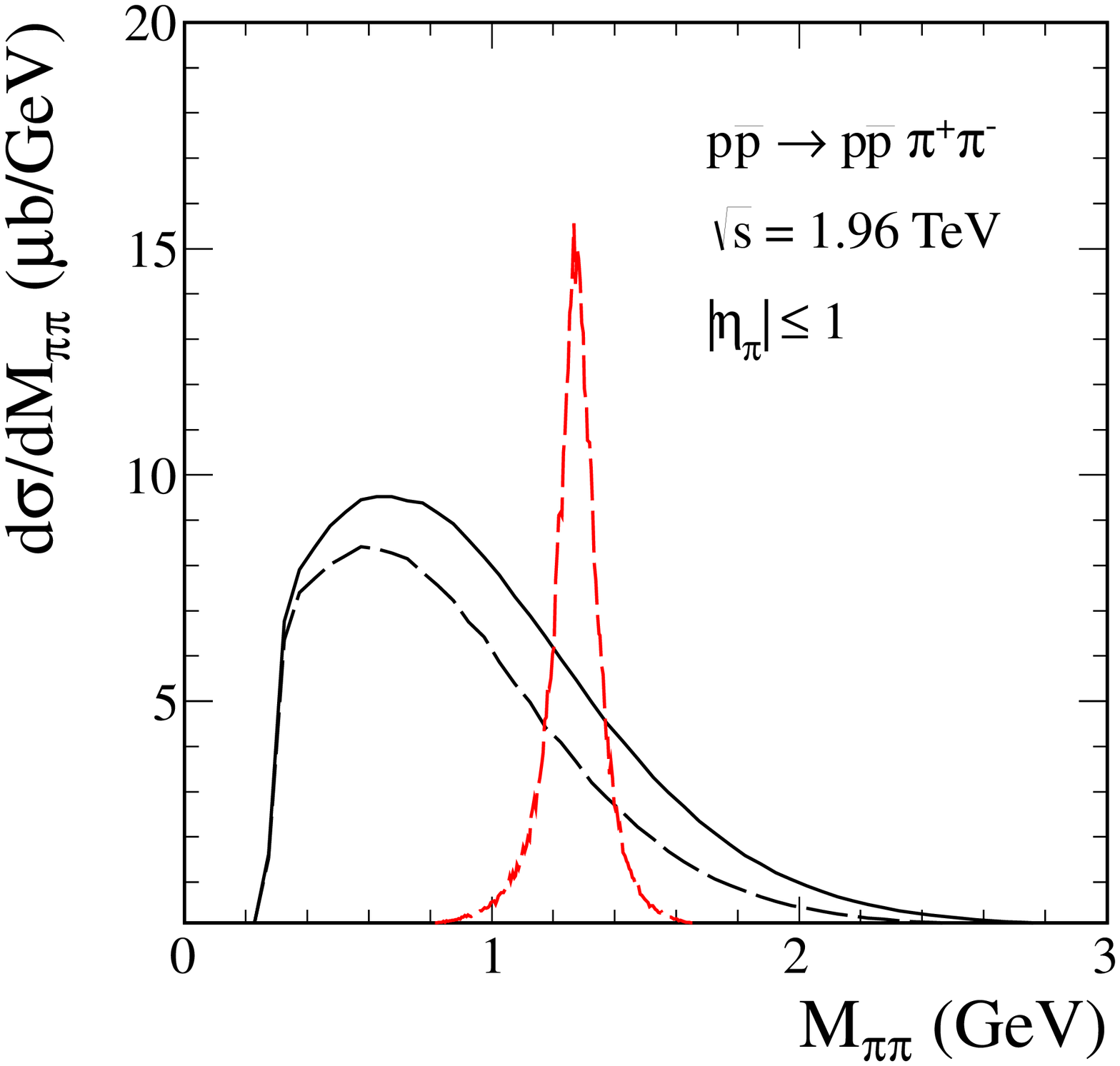} }
\caption{Left panel:
The diffractive mechanism to the central exclusive meson production.
Central panel:
Comparison of two-pion invariant mass distribution 
obtained with tensorial Pomeron exchanges
with the ISR data \cite{ABCDHW90}.
Right panel:
Predictions of $d\sigma/dM_{\pi\pi}$ distribution at $\sqrt{s} = 1.96$ TeV.
}
\label{fig:4}       
\end{figure}

Finally let us discuss results for the
$pp \to p (f_{2} \to \pi^{+} \pi^{-}) p$ reaction
calculated according to the diagram in Fig.\ref{fig:4} (left panel)
where the tensorial nature of the Pomeron
in the central meson production was used \cite{talkN}.
After the comparison of our results with cross section from existing ISR data
we present two-pion invariant mass distribution of continuum
where clear signal $f_{2}(1270)$ can be observed.
In principle, the resonance and continuum contributions should be added
coherently together leading to the distortion of the $f_{2}$
line shape as observe for the $\gamma \gamma \to f_{2}(1270) \to \pi^{+} \pi^{-}$ 
reaction \cite{SS2003}.
Correlation observables could be particularly sensitive to the spin
aspects of the Pomeron.

\section{Conclusions}
We have calculated several differential observables for the
$pp \to pp \pi^{+} \pi^{-}$ \cite{SL09,LSK09,LS10} and
$pp K^{+}K^{-}$ \cite{LS11_kaons} reactions.
The full amplitude of central diffractive process
was calculated in a simple model with parameters adjusted to low energy data.
At high energies the pions or kaons from presented CEP mechanism
are emitted preferentially
in the same hemispheres, i.e. $y_3, y_4 >$ 0 or $y_3, y_4 <$ 0.
We have predicted large cross sections for RHIC, Tevatron and LHC
which allows to hope that presented by us distributions 
will be measured in near future.
We have calculated also contributions of several diagrams
where pions/kaons are emitted from the proton lines.
These mechanisms contribute at forward and backward regions
and do not disturb the observation of the central DPE component 
which dominates at midrapidities.

At the Tevatron the measurement of exclusive production of $\chi_{c}$
via decay in the $J/\psi + \gamma$ channel cannot provide production
cross sections for different species of $\chi_{c}$.
In this decay channel the contributions of $\chi_{c}$ mesons
with different spins are similar and experimental resolution is not
sufficient to distinguish them. At the LHC situation should be better.
We have analyzed a possibility to measure the
exclusive production of $\chi_{c0}$ meson in the proton-(anti)proton
collisions at the LHC, Tevatron and RHIC 
via $\chi_{c0} \to \pi \pi$, $KK$ decay channels.
We find that the relative contribution of resonance states
and dipion/dikaon continuum strongly depend on the cut on pion/kaon $p_{t}$.
The cuts play then a role of the $\pi \pi$ or $KK$ resonance filter.
We demonstrated how to impose extra cuts in order to
improve the signal-to-background ratio.
For a more detailed discussion of this issue see \cite{LPS11,LS11_kaons}.

We have found that tensorial Pomeron may equally well describe
experimental data on exclusive meson production as the vectorial Pomeron
used in the literature.
Future experimental data on exclusive meson production at higher
energies may give a better answer on the spin structure of the Pomeron
and its coupling to the nucleon and mesons. 
We have shown that relevant measurements at high energies are possible
and could provide useful information 
e.g. about the $f_{0}$, $f_{2}$ exclusive production.
\\

PL thanks the Organisers for providing excellent 
scientific environment during the Conference.
This work was supported in part by the MNiSW grant
No. PRO-2011/01/N/ST2/04116.


\begin{thebibliography}{}

\bibitem{COMPASS}
F. Nerling, for the COMPASS Collaboration, talk at this Conference (Meson2012);
A. Austregesilo and T. Schlueter, arXiv:1207.0949.

\bibitem{Turnau}
J. Turnau, for the STAR Collaboration, talk at this Conference (Meson2012).

\bibitem{Zurek}
M. Albrow, A. Swiech and M. Zurek, talk at this Conference (Meson2012).

\bibitem{LPS11}
P. Lebiedowicz, R. Pasechnik and A. Szczurek, Phys. Lett. {\bf B701} (2011) 434.

\bibitem{Lang_Khoze}
L.A. Harland-Lang, V.A. Khoze, M.G. Ryskin and W.J. Stirling, arXiv:1204.4803;
arXiv:1105.1626.

\bibitem{SL09}
A. Szczurek and P. Lebiedowicz, Nucl. Phys. \textbf{A826} (2009) 101.

\bibitem{LSK09}
P. Lebiedowicz, A. Szczurek and R. Kami\'nski, Phys. Lett. \textbf{B680} (2009) 459.

\bibitem{LS10}
P. Lebiedowicz and A. Szczurek, Phys. Rev. \textbf{D81} (2010) 036003.

\bibitem{LS11_kaons}
P. Lebiedowicz and A. Szczurek, Phys. Rev. \textbf{D85} (2012) 014026.

\bibitem{CDF}
T. Aaltonen \textit{et al.} [CDF Collaboration], Phys. Rev. Lett. \textbf{102} (2009) 242001.

\bibitem{PST_chic}
R.S. Pasechnik, A. Szczurek and O.V. Teryaev, Phys. Rev. {\bf D78} (2008) 014007;
Phys. Lett. {\bf B680} (2009) 62; Phys. Rev. {\bf D81} (2010) 034024.

\bibitem{LKRS10}
L.A. Harland-Lang, V.A. Khoze, M.G. Ryskin and W.J. Stirling, Eur. Phys. J. {\bf C65} (2010) 433.

\bibitem{SLTCS11}
R. Staszewski, P. Lebiedowicz, M. Trzebi\'nski, J. Chwastowski and A. Szczurek,
Acta Phys. Polon. \textbf{B42} (2011) 1861.

\bibitem{LS11}
P. Lebiedowicz and A. Szczurek, Phys. Rev. \textbf{D83} (2011) 076002.

\bibitem{ABCDHW89}
A. Breakstone \textit{et al.} [ABCDHW Collaboration], Z. Phys. {\bf C42} (1989) 387.

\bibitem{ABCDHW90}
A. Breakstone \textit{et al.} [ABCDHW Collaboration], Z. Phys. {\bf C48} (1990) 569.

\bibitem{GJR}
M. Gl\"uck, D. Jimenez-Delgado, E. Reya and C. Schuck, Phys. Lett. {\bf B664} (2008) 133.

\bibitem{GRV}
M. Gl\"uck, E. Reya and A. Vogt, Z. Phys. {\bf C67} (1995) 433.

\bibitem{talkN}
O. Nachtmann, talk "A model for high-energy soft reactions" at ECT* workshop on
Exclusive and diffractive processes in high energy proton-proton
and nucleus-nucleus collisions, Trento, February 27 - March 2, 2012.

\bibitem{SS2003}
A. Szczurek and J. Speth, Nucl. Phys. {\bf A 728} (2003) 182.

\end{thebibliography}
\end{document}